\documentclass[12pt]{article}

\usepackage{amsfonts}
\usepackage{amssymb}
\usepackage{amsthm}
\usepackage[fleqn]{amsmath}
\usepackage[mathcal]{euscript}
\usepackage{mathrsfs}

\newcommand{\Ref}[1]{(\ref{#1})}

\newcommand{\kB}{k_{\mathrm{B}}} 

\newcommand{\veps}{\varepsilon}

\newcommand{\vect}[1] {\boldsymbol{{ #1}} }

\newcommand{\pV}{{\vect{p}}}           
\newcommand{\qV}{{\vect{q}}}           

\DeclareMathAlphabet{\mathpzc}{OT1}{pzc}{m}{it}

\newcommand{\abs}[1]{\left| #1 \right|}

\newcommand{\dd}{\mathrm{d}}

\newcommand{\Rset}{\mathbb{R}}
\newcommand{\Sset}{\mathbb{S}}

\newcommand{\cE}{{\cal E}}

\newcommand{\tfrhalf}{{\textstyle{\frac{1}{2}}}}

\begin{document}

\title{On Ruelle's construction of the thermodynamic limit for the classical microcanonical entropy}

\vspace{-0.3cm}
\author{\normalsize \sc{Michael K.-H. Kiessling}\\[-0.1cm]
	\normalsize Department of Mathematics, Rutgers University\\[-0.1cm]
	\normalsize Piscataway NJ 08854, USA}
\vspace{-0.5cm}
\date{$\phantom{nix}$}
\maketitle
\vspace{-1.8cm}

\begin{abstract}
\noindent
	In 1969 Ruelle published his construction of the thermodynamic limit, in the sense of Fisher, 
for the quasi-microcanonical entropy density of classical Hamiltonian $N$-body systems with stable and 
tempered pair interactions. 
	Here, ``quasi-microcanonical'' refers to the fact that he discussed the entropy defined
with a regularized microcanonical measure as
$\ln(N!^{-1}\int\chi_{\left\{\cE-\triangle \cE < H <\cE\right\}}\dd^{6N}\!X)$  
rather than defined with the proper microcanonical measure as
$\ln(N!^{-1}\int\delta(\cE-H)\;\dd^{6N}\!X)$.
	Replacing $\delta(\cE-H)$ by $\chi_{\left\{\cE-\triangle \cE < H <\cE\right\}}$ 
seems to have become the standard procedure for rigorous treatments of the microcanonical ensemble hence. 
	In this note we make a very elementary technical observation to the effect that Ruelle's proof
(still based on regularization) does establish the thermodynamic limit also for the entropy density 
defined with the proper microcanonical measure.
	We also show that with only minor changes in the proof the regularization of 
$\delta(\cE-H)$ is actually not needed at all.

\end{abstract}
\noindent
The object of interest in this note is Boltzmann's ergodic ensemble entropy
\begin{equation}
S_{H^{(N)}_\Lambda}(\cE) 
= \ln \Omega^\prime_{H^{(N)}_\Lambda}(\cE),
\label{eq:BOLTZMANNentropy}
\end{equation}
\vskip-.2truecm
\noindent
where
\begin{equation}
\Omega^\prime_{H^{(N)}_\Lambda}(\cE) 
=
 {\textstyle{\frac{1}{N!}}}\int \delta\!\left(\cE-H^{(N)}_\Lambda(X^{(N)})\right)\dd^{6N}\!X
\label{eq:STRUKTURfunktion}
\end{equation}
\vskip-.2truecm
\noindent
is known as the \textit{structure function}; here, the $^\prime$ means derivative w.r.t. $\cE$ of
\begin{equation}
\Omega_{H^{(N)}_\Lambda}(\cE) 
= 
 {\textstyle{\frac{1}{N!}}}\int \chi_{\left\{H^{(N)}_\Lambda <\cE\right\}}\dd^{6N}\!X\,,
\label{eq:STRUKTURfunktionPRIMITIVE}
\end{equation}
\vskip-.2truecm
\noindent
where $\dd^{6N}\!X:= \dd^{3N}\!p\,\dd^{3N}\!q$ and $\chi_{\{H^{(N)}_\Lambda<\cE\}}$ is the characteristic 
function of the set $\{H^{(N)}_\Lambda(X^{(N)})<\cE\}\!\subset\Rset^{3N}\!\times\Lambda^{N}$, 
with $X^{(N)}\!: =(\pV_1,...,\pV_N;\qV_1,...,\qV_N)\!\in \Rset^{3N}\!\times\Lambda^{N}$,  
\smallskip
\hrule
\smallskip\noindent
{\small 
Typeset in \LaTeX\ by the author. Original version: August 08, 2008. Revised: October 11, 2008.
To appear in Journal of Statistical Physics.

\noindent
\copyright 2008 The author. 
This preprint may be reproduced for noncommercial purposes.}
\newpage

\noindent
and where
\begin{equation} 
H^{(N)}_\Lambda (X^{(N)})
=
 \sum_{1\leq i\leq N}
{\tfrhalf}\abs{\pV_i}^2
+
\sum\sum_{\hskip-.7truecm 1\leq i < j\leq N} W(|\qV_i-\qV_j|)
+
 \sum_{1\leq i\leq N} V_\Lambda(\qV_i)
\label{eq:HAM}
\end{equation}	
\vskip-.2truecm
\noindent
is the  Hamiltonian 
of a Newtonian single specie $N$-body system in $\Lambda\subset\Rset^3$.\footnote{We use units of 
		$\kB$ for entropy, $mc^2$ for energy,  $mc$ for momentum, $h/mc$ for length, where $m$ is 
		the particle mass, $c$ the speed of light, $\kB$ Boltzmann's and $h$ Planck's constant.}
	The entropy \Ref{eq:BOLTZMANNentropy} can be evaluated in 
great detail for the perfect gas Hamiltonian ($W\equiv 0$) \cite{Boltzmann}. 
	However, for $W{\not\equiv}0$ an exact evaluation would seem virtually impossible\footnote{Hard 
		sphere interactions merit special mention because they allow one to compute at least
		the $\cE$-dependence of \Ref{eq:BOLTZMANNentropy} exactly (as for the perfect gas).}
and one has to resort to asymptotic analysis for large $N$ \cite{MazurvdLinden}.

	Ideally one wishes to show that the entropy density $|\Lambda|^{-1}S_{H^{(N)}_\Lambda}(\cE)$ converges 
in the \emph{thermodynamic limit} where $N\to\infty$ such that $\Lambda$ grows ``evenly'' with $N$ (in the 
sense of Fisher \cite{Fisher}), and such that $N/|\Lambda| \to \rho$ with $\rho\in\Rset_+$ fixed, and 
$\cE/|\Lambda|\to\veps$ with $\veps\in\Rset$ fixed; furthermore, the limit function $s(\rho,\veps)$ 
should have the right thermodynamic properties. 
	To avoid a trivial thermodynamic limit (negative infinite entropy per volume) where all particles either 
end up ``at infinity'' or else all coalesce to a point, the \textit{configurational Hamiltonian}
\begin{equation}
U_\Lambda^{(N)}(\qV_1,...,\qV_N)
=
\sum\sum_{\hskip-.7truecm 1\leq i < j\leq N} W(|\qV_i-\qV_j|)
+
 \sum_{1\leq i\leq N} V_\Lambda(\qV_i)
\label{eq:KHAM}
\end{equation}
\vskip-.2truecm
\noindent
is assumed to be stable (bounded below $\propto N$) and tempered (``short range''; for the precise
definition, see \cite{Fisher} and \cite{RuelleBOOK}).
	The r\^{o}le of the single particle potential $V_\Lambda$ is merely to confine the $N$ particles
to the domain $\Lambda$, so one can take $V_\Lambda(\qV)=+\infty$ whenever $\qV\not\in\Lambda$, and $V_\Lambda(\qV) =0$ else.
	Pair potentials $W$ of interested in chemical and condensed matter physics, such as hard sphere 
or Lennard-Jones interactions, satisfy the postulated conditions on $U_\Lambda^{(N)}$.\footnote{Like 
		Ruelle \cite{RuelleBOOK}, in \Ref{eq:KHAM} we could also allow irreducible higher-order 
		many-body interactions which are permutation symmetric, translation-invariant, stable and tempered.}

	In chapter 3 of his book \cite{RuelleBOOK}, Ruelle proved that under the above mentioned conditions,
when $\veps$ and $\rho$ are fixed in an admissible joint domain $\Theta$, then the thermodynamic limit of the 
\emph{quasi-microcanonical} ensemble entropy 
\begin{equation}
S^{-}_{H^{(N)}_\Lambda}(\cE) = \ln \Omega_{H^{(N)}_\Lambda}(\cE),
\label{eq:RUELLEentropyB}
\end{equation}
taken per volume, exists and is a concave continuous increasing function 
$s_{\mathrm{tot}}(\rho,\veps)$ on $\Theta$. 
	He also showed the same result obtains if $\Omega_{H^{(N)}_\Lambda}(\cE)$ in \Ref{eq:RUELLEentropyB} 
is replaced by $\Omega_{H^{(N)}_\Lambda}(\cE) -\Omega_{H^{(N)}_\Lambda}(\cE-\triangle \cE)$;
however, \emph{this} argument works only for $\triangle \cE >0$ and does not capture \Ref{eq:BOLTZMANNentropy}.

	Replacing the microcanonical measure $\delta\bigl(\cE-H^{(N)}_\Lambda(X^{(N)})\bigr)$
by a quasi-micro\-canonical measure $\chi_{\{\cE-\triangle \cE < H^{(N)}_\Lambda<\cE\}}$
or $\chi_{\{H^{(N)}_\Lambda<\cE\}}$ goes back at least to \cite{Gibbs} and
seems to have become the standard procedure for rigorous 
treatments of the microcanonical ensemble \cite{RuelleBOOK,Lanford,MartinLoef}.
	The purpose of this brief note is to point out that no regularization of the classical microcanonical
measure is necessary, and actually never was.
	We first make an elementary technical observation which shows that Ruelle's proof basically establishes
the thermodynamic limit for Boltzmann's ergodic ensemble entropy \Ref{eq:BOLTZMANNentropy} per volume; a key formula 
in this proof is still based on the regularized measures.
	A minor variation on the theme of Ruelle's proof finally shows that the regularization is not needed.


	A key ingredient in Ruelle's proof is the reduction of the $(\pV,\qV)$-space problem to two separate problems, one
in $\pV$-space and the other in $\qV$-space. 
	Namely, since the characteristic function of an interval of $\Rset$ is a non-negative, bounded, piecewise continuous 
function, it is the upper limit of a sequence of continuous functions, and as such weakly lower semi-continuous.
	Therefore the convolution integral with a $\delta$ function is well-defined and yields the identity
\begin{equation}
	\chi_{\left\{H^{(N)}_\Lambda<\cE\right\}}
=
\int\chi_{\left\{U^{(N)}_\Lambda<\cE - E \right\}}\delta \left(E-  K^{(N)} \right)\dd{E}, 
\label{eq:chideltaCONVOLUTION}
\end{equation}
where we introduced the abbreviation $K^{(N)}$ for the kinetic Hamiltonian, i.e.
\begin{equation}
 K^{(N)}  (\pV_1,...,\pV_N) = {\textstyle{\sum_{i=1}^N}}\, {\tfrhalf}\abs{\pV_i}^2 ,
\label{eq:kinENERGY}
\end{equation}
so $H^{(N)}_\Lambda = K^{(N)}+ U^{(N)}_\Lambda$. 
	Integrating \Ref{eq:chideltaCONVOLUTION} w.r.t. $\dd^{6N}X$ and interchanging with the $\dd{E}$ 
integration on the so integrated r.h.s.\Ref{eq:chideltaCONVOLUTION}, then
multiplying by $N!^{-1}$, yields Ruelle's eq.(4.3) in sect. 3.4 of \cite{RuelleBOOK},
\begin{equation}
\Omega_{H^{(N)}_\Lambda}(\cE)
= 
\int\Omega_{U^{(N)}_\Lambda}(\cE - E)\;\Omega^\prime_{K^{(N)}}(E)\,\dd{E} ,
\label{eq:PRIMofSTRUCTUREfctDECOMP}
\end{equation}
\vskip-.2truecm
\noindent
where
\begin{equation}
\Omega_{U^{(N)}_\Lambda}(\cE) 
= 
 {\textstyle{\frac{1}{N!}}}\int\!\!\chi_{\left\{U^{(N)}_\Lambda<\cE\right\}} \dd^{3N}\!q
\label{eq:RUELLEconfigINTEGRAL}
\end{equation}
\vskip-.2truecm
\noindent
and
\begin{equation}
\Omega^\prime_{K^{(N)}}(\cE)
= 
\int \delta \left(\cE-  K^{(N)} \right)\dd^{3N}p ;
\label{eq:kineticSTRUCTUREfct}
\end{equation}
\vskip-.2truecm
\noindent
note that multiplying $\Omega^\prime_{K^{(N)}}(\cE)$ by a factor $|\Lambda|^N/N!$ gives the structure function of the
perfect gas.
	Thanks to \Ref{eq:PRIMofSTRUCTUREfctDECOMP}, the proof that the thermodynamic limit exists for the logarithm 
of l.h.s.\Ref{eq:PRIMofSTRUCTUREfctDECOMP}, taken per volume, reduces to proving 
that the thermodynamic limit exists separately for the logarithm of \Ref{eq:RUELLEconfigINTEGRAL} and
of \Ref{eq:kineticSTRUCTUREfct}, each taken per volume.
	For \Ref{eq:kineticSTRUCTUREfct} this is easy.
	The $\pV$-space integrations in \Ref{eq:kineticSTRUCTUREfct} can be carried out explicitly 
as for the perfect gas, yielding
\begin{equation}
\Omega^\prime_{K^{(N)}}(\cE)
= 
{\textstyle{\frac{(2\pi)^{3N/2}}{\Gamma(3N/2)}}}\cE^{\frac{3N}{2} -1}\chi_{_{\{\cE>0\}}},
\label{eq:kineticSTRUCTUREfctEXPL}
\end{equation}
\vskip-.1truecm
\noindent
and so one can take the thermodynamic limit of $|\Lambda|^{-1}\ln  \Omega^\prime_{K^{(N)}}(\cE)$, giving
\begin{equation}
\lim_{N\to\infty} {\textstyle{\frac{1}{|\Lambda|}}} \ln \Omega^\prime_{K^{(N)}}(\cE) 
=
{\textstyle{\frac{3}{2}}}\rho \ln\bigl({\textstyle{\frac{4\pi e}{3}\frac{\veps}{\rho}}} \bigr)
\equiv
 s_{\mathrm{kin}}(\rho,\veps)
\label{eq:kinENTROPYlim}
\end{equation}
\vskip-.15truecm
\noindent
(cf., eq.(4.4) in sect.3.4 of \cite{RuelleBOOK}),
which differs from the entropy density of the perfect gas by an added $\rho\ln(\rho/e)$, due
to the absence of the factor $|\Lambda|^N/N!$ in \Ref{eq:kineticSTRUCTUREfct}.
	All the hard technical work, which we won't repeat here (and don't need to), 
now goes into analyzing the configurational integral \Ref{eq:RUELLEconfigINTEGRAL}.
	Ruelle proves (Thm. 3.3.12): \emph{If $\Lambda\to\Rset^3$ (Fisher) when $N\to\infty$, 
such that $\cE/ |\Lambda|\to \veps$ and $N/|\Lambda|\to \rho$ with $\veps$ and $\rho$ fixed in an 
admissible joint domain $\Theta$, then the limit for the configurational (interaction) entropy density exists,
\begin{equation}
\lim_{N\to\infty} {\textstyle{\frac{1}{|\Lambda|}}} \ln \Omega_{U^{(N)}_\Lambda}(\cE) 
=
 s_{\mathrm{int}}(\rho,\veps),
\label{eq:configENTROPYlim}
\end{equation}
\vskip-.1truecm
\noindent
and $s_{\mathrm{int}}(\rho,\veps)$ is concave and continuous on $\Theta$.}
	Having \Ref{eq:kinENTROPYlim} and \Ref{eq:configENTROPYlim} Ruelle
now applies Laplace's method\footnote{For background material on this
		method, cf. \cite{Ellis}, sect. II.7.}
to \Ref{eq:PRIMofSTRUCTUREfctDECOMP} and finds (subsect. 3.4.1 and 3.4.2)
\begin{equation}
\lim_{N\to\infty} {\textstyle{\frac{1}{|\Lambda|}}} \ln \Omega_{H^{(N)}_\Lambda}(\cE) 
= 
\sup_{\tilde{\veps}\in (0,\veps -\veps_0(\rho))}
    \lbrace s_{\mathrm{kin}}(\rho,\tilde{\veps}) + s_{\mathrm{int}}(\rho,\veps-\tilde{\veps})\rbrace
\equiv
 s_{\mathrm{tot}}(\rho,\veps)
,
\label{eq:RuelleVP}
\end{equation}
\vskip-.2truecm
\noindent
where $\veps_0(\rho) = \inf\{\pi_2(\rho,\veps)|\rho\ {\mathrm{fixed}}\}$ is a boundary point of $\Theta$.
	The continuity, concavity, and increase of $ s_{\mathrm{tot}}(\rho,\veps)$ on $\Theta$ follow from
\Ref{eq:RuelleVP}.
	Ruelle also shows (see subsect. 3.3.14) that $s_{\mathrm{int}}(\rho,\veps)$ remains unchanged if
in \Ref{eq:configENTROPYlim} one  replaces $\Omega_{U^{(N)}_\Lambda}(\cE)$ by 
$\Omega_{U^{(N)}_\Lambda}(\cE) -\Omega_{U^{(N)}_\Lambda}(\cE-\triangle\cE)$ with $\triangle\cE>0$.
	It follows that the same is also true for $s_{\mathrm{tot}}(\rho,\veps)$ in \Ref{eq:RuelleVP}.
	This completes our summary of Ruelle's proof of the thermodynamic limit of the quasi-microcanonical 
entropy per volume.

	Interestingly enough, by taking the derivative w.r.t. $\cE$ of 
the representation \Ref{eq:PRIMofSTRUCTUREfctDECOMP}, one obtains Gibbs' eq.(303) \cite{Gibbs},
 \begin{equation}
\hskip-.2truecm
\Omega^\prime_{H^{(N)}_\Lambda}(\cE)
= 
\int\Omega^\prime_{U^{(N)}_\Lambda}(\cE - E)\;\Omega^\prime_{K^{(N)}}(E)\,\dd{E},
\label{eq:STRUCTUREfctDECOMPalaGIBBS}
\end{equation}
\vskip-.2truecm
\noindent
with $\Omega^\prime{}_{U^{(N)}_\Lambda}(\cE)$ generally defined only in the distributional sense.
	But exchanging $\cE$ and $E$ derivatives under the integral in \Ref{eq:STRUCTUREfctDECOMPalaGIBBS},
integrating by parts, then taking logarithms, now yields the following representation for 
\Ref{eq:BOLTZMANNentropy}:
\begin{equation}
S_{H^{(N)}_\Lambda}(\cE) 
= 
\ln \int\Omega_{U^{(N)}_\Lambda}(\cE - E)\;\Omega^{\prime\prime}_{K^{(N)}}(E)\,\dd{E}.
\label{eq:lnSTRUCTUREfctDECOMP}
\end{equation}
\vskip-.2truecm
\noindent
	Next, \Ref{eq:kineticSTRUCTUREfctEXPL} shows that in \Ref{eq:kinENTROPYlim} 
we can replace $\Omega^{\prime}_{K^{(N)}}(\cE)$
by $\Omega^{\prime\prime}_{K^{(N)}}(\cE)$ and still get $s_{\mathrm{kin}}(\rho,\veps)$,
and this fact plus the limit \Ref{eq:configENTROPYlim} plus an easy adaptation of the Laplace method arguments
in subsect. 3.4.1 and 3.4.2 of \cite{RuelleBOOK} now prove:\hfill 

\noindent
\emph{The thermodynamic limit for Boltzmann's 
entropy per volume exists and has all the right monotonicity, continuity and concavity properties. 
	This limit coincides with Ruelle's $s_{\mathrm{tot}}(\rho,\veps)$ given by the variational 
principle \Ref{eq:RuelleVP}, thus} 
\begin{equation}
\lim_{N\to\infty} {\textstyle{\frac{1}{|\Lambda|}}} \ln \Omega^\prime_{H^{(N)}_\Lambda}(\cE) 
=
 s_{\mathrm{tot}}(\rho,\veps).
\label{eq:BoltzmannENTROPYlim}
\end{equation}
\vskip-.1truecm

	This concludes our demonstration that Ruelle's treatment of \Ref{eq:RUELLEentropyB} 
basically achieves control over \Ref{eq:BOLTZMANNentropy}.
	This entirely elementary observation may well have been made before; yet the author is not aware of
anyone having pointed it out.

	Armed with hindsight, we now inquire into whether \Ref{eq:lnSTRUCTUREfctDECOMP} can be obtained directly
from \Ref{eq:BOLTZMANNentropy}, i.e. without first proving \Ref{eq:PRIMofSTRUCTUREfctDECOMP} for the regularized 
ensemble entropy \Ref{eq:RUELLEentropyB} and then taking the derivative of \Ref{eq:PRIMofSTRUCTUREfctDECOMP}.
	At the  purely formal level this is quite straightforward.
	Rather than from \Ref{eq:chideltaCONVOLUTION} we start from the ``identity''
\begin{equation}
	\delta\left(\cE -H^{(N)}_\Lambda\right)
=
{\phantom{\Big(}}^{``}\int
	\delta\bigl(\cE-E-U^{(N)}_\Lambda \bigr)\delta \left(E-K^{(N)} \right)\dd{E},
{\phantom{\Big)}}^{^{\!\!\!,,}}
\label{eq:deltadeltaCONVOLUTION}
\end{equation}
\vskip-.2truecm
\noindent
which happens to be the formal derivative w.r.t. $\cE$ of the identity \Ref{eq:chideltaCONVOLUTION}; 
we now formally integrate by parts on the r.h.s.\Ref{eq:deltadeltaCONVOLUTION} to obtain the ``identity''
\begin{equation}
	\delta\left(\cE -H^{(N)}_\Lambda\right)
=
{\phantom{\Big(}}^{``}\int
	\chi_{_{\left\{U^{(N)}_\Lambda <\cE-E\right\}}}\delta^\prime \left(E-K^{(N)} \right)\dd{E};
{\phantom{\Big)}}^{^{\!\!\!,,}}
\label{eq:chideltaprimeCONVOLUTION}
\end{equation}
\vskip-.2truecm
\noindent
next we integrate \Ref{eq:chideltaprimeCONVOLUTION} w.r.t. $\dd^{6N}\!X$, and in the so integrated 
r.h.s.\Ref{eq:chideltaprimeCONVOLUTION} we formally interchange  $\dd^{6N}\!X$ and $\dd{E}$ integrations, 
multiply by $N!^{-1}$, then take logarithms, et voil\`a: out pops \Ref{eq:lnSTRUCTUREfctDECOMP}.
	Unfortunately, these are all only symbolic manipulations.\footnote{However, one can
		see why the ease with which such formal manipulations apparently lead to the correct
		result does make it desirable to seek their rigorous foundation \cite{Colombeau}.}
	A slightly different plan of attack leads to conquest, though.

	We note that the $\pV$-space integrations involved in \Ref{eq:BOLTZMANNentropy} can be carried out 
in the same fashion as for the perfect gas.
	The problem then becomes to study the large $N$ asymptotics of the resulting $\qV$-space
integrals.
	Thus, carrying out the $\pV$ integrations in $\Omega^\prime_{H^{(N)}_\Lambda}(\cE)$ given 
by \Ref{eq:STRUKTURfunktion}, with $H^{(N)}_\Lambda$ given by \Ref{eq:HAM}, Boltzmann's
\vskip-.15truecm
\noindent
entropy \Ref{eq:BOLTZMANNentropy} becomes\quad (cf. eq.(305) in \cite{Gibbs})
\begin{equation}
\hskip-.2truecm
S_{H^{(N)}_\Lambda}(\cE) 
= 
\ln\!\left({\textstyle{\frac{2^{3N/2}}{3N}}}\left|\Sset^{3N-1}\right|\!\Psi^\prime_{U^{(N)}_\Lambda}(\cE) \right)
\label{eq:MCentropyINTEGRATEDinP}
\end{equation}
\vskip-.2truecm
\noindent
with
\begin{equation}
\Psi^\prime_{U^{(N)}_\Lambda}(\cE) 
=
{\textstyle{\frac{3/2}{(N-1)!}}}
\int\!\!\!
	\left(\cE-U_\Lambda^{(N)}(\qV_1,...,\qV_N)\right)^{\frac{3N}{2}-1}\!\chi_{\left\{U^{(N)}_\Lambda<\cE\right\}}
 \dd^{3N}\!q.
\label{eq:MCconfigINTEGRAL}
\end{equation}
\vskip-.2truecm
\noindent
	The primitive of \Ref{eq:MCconfigINTEGRAL}, 
\begin{equation}
\Psi_{U^{(N)}_\Lambda}(\cE) 
= 
{\textstyle{\frac{1}{N!}}}
\int\!\!\!
	\left(\cE-U_\Lambda^{(N)}(\qV_1,...,\qV_N)\right)^{\frac{3N}{2}}\!\chi_{\left\{U^{(N)}_\Lambda<\cE\right\}}
 \dd^{3N}\!q,
\label{eq:MCconfigINTEGRALprimitive}
\end{equation}
\vskip-.2truecm
\noindent
in turn obtains when carrying out the $\pV$ integrations in  $\Omega_{H^{(N)}_\Lambda}(\cE)$ given in
\Ref{eq:STRUKTURfunktionPRIMITIVE}, so that the entropy \Ref{eq:RUELLEentropyB} reads (cf. eq.(304) in \cite{Gibbs})
\begin{equation}
S^{-}_{H^{(N)}_\Lambda}(\cE) 
= 
\ln\Big({\textstyle{\frac{2^{3N/2}}{3N}}}\left|\Sset^{3N-1}\right|\!\Psi_{U^{(N)}_\Lambda}(\cE)\Big).
\label{eq:RUELLEentropyBasLogPsi}
\end{equation}
\vskip-.2truecm
\noindent
	The integral \Ref{eq:MCconfigINTEGRAL} is obviously not more complicated than \Ref{eq:MCconfigINTEGRALprimitive}.
	Since Ruelle controlled \Ref{eq:RUELLEentropyB}, which is \Ref{eq:RUELLEentropyBasLogPsi}, he effectively
controlled \Ref{eq:MCconfigINTEGRALprimitive}, and this means that his arguments control \Ref{eq:MCconfigINTEGRAL}, the 
miniscule difference between the powers ${3N}/{2}$ and $({3N}/{2})-1$ at the integrands of 
\Ref{eq:MCconfigINTEGRALprimitive} and \Ref{eq:MCconfigINTEGRAL}, and between their factors $1/N!$ and $(3/2)/(N-1)!$
hardly making a difference at all.

	All we need to do is to show how Ruelle's control of $\Omega_{U^{(N)}_\Lambda}(\cE)$ given in 
\Ref{eq:RUELLEconfigINTEGRAL} implies the control of $\Psi_{U^{(N)}_\Lambda}(\cE)$ given in
\Ref{eq:MCconfigINTEGRALprimitive}, and this will pave the way for the control of
$\Psi^\prime_{U^{(N)}_\Lambda}(\cE)$ given in \Ref{eq:MCconfigINTEGRAL}.
	In the spirit of Ruelle's proof, we seek a convolution representation of 
$\Psi_{U^{(N)}_\Lambda}(\cE)$ and $\Psi^\prime_{U^{(N)}_\Lambda}(\cE)$ involving $\Omega_{U^{(N)}_\Lambda}(\cE)$.
	Through integration by parts and Fubini's theorem one easily verifies that
\begin{equation}
\int\!\!\! \left(\cE-U_\Lambda^{(N)}\right)^{P}\!\chi_{\left\{U^{(N)}_\Lambda<\cE\right\}} \dd^{3N}\!q
= 
P\int_0^{\cE-\cE_g} E^{P-1} \int\!\!\chi_{\left\{U^{(N)}_\Lambda<\cE-E\right\}} \dd^{3N}\!q\,\dd{E}
\label{eq:PSIconfigINTEGRALconvolutionP}
\end{equation}
for any power $P>0$, and so the desired convolutions read
\begin{equation}
\Psi_{U^{(N)}_\Lambda}(\cE) 
= 
{\textstyle{\frac{3N}{2}}}
\int_0^{\cE-\cE_g}  \Omega_{U^{(N)}_\Lambda}(\cE-E)\, E^{\frac{3N}{2}-1}\dd{E},
\label{eq:PSIconfigINTEGRALconvolution}
\end{equation}
respectively
\begin{equation}
\Psi^\prime_{U^{(N)}_\Lambda}(\cE) 
= 
{\textstyle{\frac{3N}{2}}}\left({\textstyle{\frac{3N}{2}-1}}\right)
\int_0^{\cE-\cE_g} \Omega_{U^{(N)}_\Lambda}(\cE-E)\, E^{\frac{3N}{2}-2} \dd{E}.
\label{eq:PsiPRIMEconfigINTEGRALconvolution}
\end{equation}
	Inserting \Ref{eq:PSIconfigINTEGRALconvolution} into \Ref{eq:RUELLEentropyBasLogPsi} and recalling
\Ref{eq:kineticSTRUCTUREfctEXPL} gives us the log of r.h.s.\Ref{eq:PRIMofSTRUCTUREfctDECOMP}, and we are back 
full circle to the last stage of Ruelle's proof.
	In the same vein, inserting \Ref{eq:PsiPRIMEconfigINTEGRALconvolution} into 
\Ref{eq:MCentropyINTEGRATEDinP}, and again recalling \Ref{eq:kineticSTRUCTUREfctEXPL}, we get for Boltzmann's 
entropy \Ref{eq:BOLTZMANNentropy} the representation \Ref{eq:lnSTRUCTUREfctDECOMP}, and are back to our 
elementary observation made there that Ruelle's proof handles  \Ref{eq:lnSTRUCTUREfctDECOMP} as well. 

	The upshot is: the $\pV$-space integrations are already so regularizing that no 
``$\triangle \cE$ regularization'' is needed.
	Indeed, equation \Ref{eq:MCconfigINTEGRAL} 
shows explicitly that the $\pV$ integrations in $\Omega^\prime_{H^{(N)}_\Lambda}(\cE)$ given by \Ref{eq:STRUKTURfunktion}, 
with $H^{(N)}_\Lambda$ given by \Ref{eq:HAM}, automatically 
\vskip-.1truecm
\noindent
produce the characteristic function $\chi_{\{U^{(N)}_\Lambda<\cE\}}\!$ 
which is the integrand of $\Omega_{U^{(N)}_\Lambda}(\cE)$ 
\vskip-.1truecm
\noindent
given in \Ref{eq:RUELLEconfigINTEGRAL}.
	Therefore it was never necessary to replace Dirac's $\delta\bigl(\cE -H^{(N)}_\Lambda\bigr)$ 
by a characteristic function $\chi_{\{H^{(N)}_\Lambda<\cE\}}$ 
or $\chi_{\{\cE-\triangle\cE < H^{(N)}_\Lambda<\cE\}}$  in the first place.\footnote{Of course, this 
		observation is meaningless for continuous classical quasi-particle systems like point vortices 
		whose Hamiltonian is missing the kinetic $K^{(N)}(\pV_1,...,\pV_N)$ term, and 
		in which case one needs to control $\;\Omega^\prime_{U^{(N)}_\Lambda}(\cE)\;\;$\ \cite{Onsager}.
		Interestingly, for overall neutral point
		\vskip-.1truecm
\noindent
 vortex  systems this feat has been accomplished
		outside the thermodynamic limit regime, with $(\cE -\tfrhalf N\ln N)=\veps N$ scaling 
		\cite{oneilredner}, while their thermodynamic limit regime has so far been treated only with 
		the regularized $\Omega_{U^{(N)}_\Lambda}(\cE)-\Omega_{U^{(N)}_\Lambda}(\cE-\triangle\cE)$,
		see \cite{FroehlichRuelle}.}

	Needless to stress: the mathematical reasoning presented in this note has no bearing on 
quantum statistical mechanics \cite{Griffiths,RuelleBOOK}, nor on the classical statistical 
mechanics of lattice systems \cite{RuelleBOOK}.
	Both involve discrete energies, which raises not just technically but also 
conceptually different questions.
\newpage

\bigskip\noindent
\textbf{Acknowledgement:} 
I thank Sheldon Goldstein, Joel L. Lebowitz, and the two referees for their helpful comments on the manuscript.

\end{document}